\documentclass[aps,epsfig,twocolumn,showpacs,floatfix]{revtex4}
\usepackage{amsmath}
\usepackage{amssymb}
\usepackage{amsfonts}
\usepackage{epsfig}
\usepackage{graphicx}
\def\ga{\gamma}

\def\la{\lambda}
\def\th{\theta}

\def\Om{\Omega}
\newcommand{\rb} {\bar{r}}
\newcommand{\tb} {\bar{t}}
\newcommand{\rc} {\mathcal{R}}

\def\d#1#2{\displaystyle\frac{\displaystyle #1}{\displaystyle #2}}

\usepackage{color}
\def\blue{\color{blue}}

\definecolor{dyellow}{rgb}{1.,0.8,.0}
\definecolor{myblue}{rgb}{.1,.1,.7}
\definecolor{dcyan}{rgb}{.0,.6,.6}
\definecolor{dmagenta}{rgb}{0.6,0.0,0.6}
\definecolor{brown}{rgb}{0.6,0.2,0.}
\definecolor{darkblue}{rgb}{.0,.0,0.5}
\definecolor{darkred}{rgb}{0.75,0.0,0.0}
\definecolor{orange}{rgb}{1.,.6,.0}
\definecolor{dorange}{rgb}{0.8,.4,.0}
\definecolor{darkgreen}{rgb}{0.0,0.6,0.0}
\definecolor{purple}{rgb}{.4,.0,.4}

\def\bc{\begin{center}}
\def\ec{\end{center}}
\def\be{\begin{eqnarray}}
\def\ee{\end{eqnarray}}
\def\nno{\nonumber}
\newcommand{\omits}[1]{}
\begin{document}
\baselineskip=14pt
%
%PREPRINT NUMBERS
%\begin{flushright}
%\hfill
%\end{flushright}
%
%
\title{Gravitational collapse without a remnant}

\author{{Zhe Chang$^1$}} \email{changz@ihep.ac.cn}
\author{{Cheng-Bo Guan}$^{2}$} \email{guancb@sdu.edu.cn}
\author{{Chao-Guang Huang}$^1$} \email{huangcg@ihep.ac.cn}
\author{{Xin Li}$^1$} \email{lixin@ihep.ac.cn}
\affiliation{$^1$Institute of High Energy Physics, Chinese Academy
of Sciences, Beijing 100049, P. R. China} 
\affiliation{$^2$Department of Space Science and Applied Physics, Shandong University, Weihai 264209, China}

\begin{abstract}
We investigate the gravitational collapse of a spherically
symmetric, inhomogeneous star, which is described by a perfect fluid
with heat flow and satisfies the equation of state $p=\rho/3$ or
$p=C\rho^\ga$ at its center. Different from the ordinary process of
gravitational collapsing, the energy of the whole star is emitted
into space. And the remaining spacetime is a Minkowski one at the
end of the process.
\end{abstract}

\maketitle

\section{Introduction}

Gravitational collapse is one of the most important topics in
general relativity.  A large mount of models of gravitational
collapse has been proposed \cite{OS,
BOS,more,KST,choptuik,Hirschmann,SG}. In these models, the remnant
of a gravitational collapse is possible a compact self-sustained
star or a black hole, or a naked-singularity, depending on the
initial condition of the collapse.  A decade ago, in studying the
critical phenomenon in gravitational collapse, first revealed by
Choptuik \cite{choptuik}, Hirschmann and Eardley showed that the
gravitational collapse of a complex scalar field may leave behind an
approximately flat spacetime. Their model about the collapse, as a
solution of Einstein field equations, possesses self-similarity
\cite{Hirschmann}.  Later, Sch\"{a}fer and Goenner constructed a
model of gravitational contraction of a radiating spherically
symmetric body with heat flow \cite{SG}, in which both initial mass
and initial radius are infinitely large and all mass of the body can
be radiated away eventually without forming an event horizon. Also
about a decade ago, Fayos, Senovilla and Torres studied geometry
matched by two spherically symmetric spacetimes through a timelike
hypersurface from a very general point of view and exhausted all
possible and qualitatively different matchings with their
corresponding conformal diagrams for a flat Robertson-Walker model
with a linear equation of state $p= \ga \rho$ \cite{FST}.  In
particular, Fig. 9 and the time reversal of Fig. 11 in \cite{FST}
show that a flat spacetime is left when a radiating homogeneous star
(or Universe) with a linear equation of state radiates its all mass.

In the present paper, we propose a new approach to study the
gravitational collapse of a spherical star.  In our approach, a
fluid star with finite initial mass and radius is supposed to be
spherically symmetric, inhomogeneous, with the equation of state
$p=\rho/3$ or $p=C\rho^\ga$ at its center, and having heat flow
outside it.  Similar to the solutions given by Hirschmann and
Eardley\cite{Hirschmann}, Sch\"{a}fer and Goenner\cite{SG} and Fayos
et al\cite{FST},  all energy of the star in our model will be
emitted in the process of collapse, and the remaining spacetime 
is an empty flat one. It is remarkable that for a star with about a solar
mass and a solar radius, the energy at the order of $10^{54}$ {\rm
erg} will be emitted into space within about $8 \sim 600$s.  As a
result our model concludes that the inferred isotropic average
luminosities are on the order of $10^{51\sim53}$erg/s, which has the
same order of magnitude for a gamma-ray burst. 

The arrangement of the paper is as follows. The method of model
construction and the equations governing the model are described in
next section. The junction conditions and boundary conditions
are listed in Section 3. In Section 4, as examples, several
numerical solutions are presented. The concluding
remarks are given in Section 5.

\section{Formulation}  %\quad
In the pioneer paper to understand the late stages of stellar
evolutions \cite{OS}, a collapsing star is supposed to consist of
homogeneous, spherically symmetric, pressure-less, perfect fluid and
to be surrounded by an empty space.  The interior of the star may be
described by the Friedmann-Robertson-Walker metric \cite{Wein}
\be \label{FRW} ds^2=dt^2 -a^2(t)\left(\frac{dr^2}{1-kr^2} +
r^2d\Om^2\right), \ee
where $d\Om^2=d\th^2+\sin^2\th d\varphi^2$ is the metric on a unit
2-sphere, $k$ is $\frac {8\pi G}{3}$ times of the initial energy
density of dust in the unit of $c=1$, and $a(t)$ is the solution of
\be \dot a^2(t)=k[a^{-1}(t)-1]. \nno \ee

In an astrophysical environment, a star usually emits radiation and
throws out particles in the process of gravitational collapse.  In
this situation, the heat flow in the interior of a star should not
be ignored and the exterior spacetime is no longer described by a
Schwarzschild metric. To take the radiation of a star into account,
the interior solution of the gravitational collapse of radiating
stars should match to the exterior spacetime described by the Vaidya
solution \cite{Vaidya}
\be \label{Vaidya}
ds^2=(1-\frac{2GM(v)}{R})dv^2+2dv d{R}-R^2 d\Om^2 , %
\ee
which has been studied extensively \cite{BOS,more,KST,FST}.  In
particular, the gravitational collapse of a radiating spherical star
with heat flow has been studied in an isotropic coordinate system
\cite{KST}
\be ds^2=dt^2 -B^2(t,r)(dr^2+r^2d\Om^2), \ee
where
\be B(t,r)={b^2(t)}[1-\la (t) r^2]^{-1}\nno \ee
with suitable $b(t)$ and $\la(t)$ functions.  They called the
solution the Friedmann-like solution.

We study the gravitational collapse of a spherically symmetric,
inhomogeneous star in a proper-time reference, which can be written
in a generalized Friedmann coordinate system
\be \label{varyk} ds^2=dt^2 -a^2(t)\left(\frac{dr^2}{1-k(t)r^2} +
r^2d\Om^2\right) \ee
with $k$ being a function of $t$.

The stress-energy tensor of the fluid with heat flow and without
viscosity is given by
\be \label{stress} T^{\mu\nu}= (\rho+p)u^{\mu}u^{\nu}- pg^{\mu\nu}+
q^{\mu}u^{\nu}+ q^{\nu}u^{\mu}~, \ee
where $\rho$ and $p$ are the proper energy density and pressure
measured by the comoving observers respectively, $u^{\mu}$ is the
4-velocity of the fluid, and $q^{\mu}$ is the heat flow. $u^\mu$ and
$q^\mu$ satisfy
\be
u_{\mu}u^{\mu}&=&1 \\
 q_{\mu}u^{\mu}&=&0 .
\ee
For the spherically symmetric collapse, one has
\be
u^\mu &=& (u^0, u^1, 0,0) \\
q^\mu &=& (q^0, q^1, 0, 0). \ee

The Einstein's field equations
\be G_{\mu\nu}=-8\pi{G}{T_{\mu\nu}} \ee
and the covariant conservation of stress-energy tensor give rise to
\be 8\pi{G}\rho &= & \d
{k+\dot{a}^2+2a\ddot a}{a^2}+ \d Y {a^2A^2}, \label{rho}\\
8\pi{G}p &=& - \d{k+\dot{a}^2+2a\ddot{a}}{a^2} -
           \d X {a^2A^2}, \label{eq2-2}
\ee
\be \label{u0}
u_0^{\pm}&=&\sqrt{\frac{Y (X+Y) -\frac {Z^2} 2\pm Z\sqrt{\frac {Z^2} 4- X  Y} } {(X+Y)^2 -Z^2 } }, \qquad \\
\label{u1}
u_1&=& aA\sqrt{u_0^2-1}, \ee
\be \label{q0}
8\pi{G}q_0&=& \frac{1}{2a^{2}A^2}\left[\frac{Y}{u_0}-(Y-X)u_0\right],\\
8\pi{G}q_1&=& \frac{1}{2}\left[\frac{X}{u_1}-\frac
{(Y-X)u_1}{a^2A^2}\right],  \label{q1}%
\ee
with the identity
\be (8\pi G)^2 q_\mu q^\mu =- \d {Z^2-4XY}{4a^4A^4} , \label{q2} \ee
where the over-dots denote the derivatives with respect to time $t$,
and
\be
A&=& \frac {1} {\sqrt{1-k(t)r^2}},  \label{A}\\
X&=& 3a\dot{a}A\dot{A}+a^2A\ddot{A},\\
\label{Y}%
Y&=& [2(k+\dot{a}^2-a\ddot{a})A-
     a\dot{a}\dot{A}-a^2\ddot{A}]A,\qquad \\
Z&=& -\frac{4}{r}a\dot{A}. %
\ee
$X, Y, Z$ should satisfy the following solvable condition,
\be\label{Solvable} Z^2\geq 4XY. \ee
For spherical collapse, the dominant energy conditions \cite{Hawk}
require
\be
&&\rho-p\geq 0,\label{dec2}\\
&&\rho+p\geq 2\sqrt{-q_{\mu}q^{\mu}},\label{dec1} \ee which lead to
\be X+Y>|Z|\geq 0. \ee

\section{Boundary conditions}%
At the center of a star $r=0$, $\dot A =\ddot A =0$ and thus $X,
Z$ vanish. Hence,
\be
&u_0|_{r=0}=1,& \ u_1|_{r=0}=0, \\
&q_0^{}|_{r=0}=0,& \ q_1^{}|_{r=0}=0, \ee
\be \label{rho0}%
8\pi{G}\rho |_{r=0}&= &3 a^{-2}{(k+\dot{a}^2)}, \\
\label{p0} 8\pi{G}p|_{r=0} &=& - a^{-2}{(k+\dot{a}^2+2a\ddot{a})} .
\ee

The exterior solution outside the collapsing star is described by
Vaidya metric (\ref{Vaidya}). The interior solution and exterior
solution should satisfy the Darmois junction conditions on the surface of
the star \cite{Darmois, Santos},
\be %
&&R_s=ar_s~,\label{vaidya-r}\\
&&M(v)=\left . \frac{a}{2G}(k+\dot{a}^2)r^3 \right |_s~,\label{vaidya-m}\\
&&p^2_s=|q_\mu q^\mu|_s, \label{pq} %
\ee
and
\be
&&\left [dt^2-a^2A^2dr^2 \right ]_s \nno \\
&=& \left [\left(1-\d {2GM(v)}{R} \right) dv^2+2 dv dR \right ]_s,
\ee
where the subscript $s$ denotes the value taken on the surface of
the star. In particular, in Eq.(\ref{pq}) with the expressions
(\ref{eq2-2}), (\ref{q2}), (\ref{A}), $r$ should be replaced by
$r_s(t)$, for Eq.(\ref{pq}) is only valid on the surface of the
star.  $M(v)$ is the Misner-Sharp mass \cite{Misner-Sharp} of the
star. (Appendix A gives the detailed discussion on the junction
condition.) The equation of motion of the surface is
\be
&&\frac{dr_s}{dt}= \left .-a^{-2}A^{-2}\frac{u_1}{u_0}\right
|_s.\label{rdot} \ee

In brief, with the help of the equation of state of fluid at the
center of the star, Eqs.(\ref{rho0}), (\ref{p0}), (\ref{pq}) and
(\ref{rdot}) constitute a complete system of equations for $k(t)$,
$a(t)$, $r_s(t)$.  Once $k(t)$, $a(t)$ and $r_s(t)$ are obtained,
one may directly get the expression for $\rho(r,t)$ and $p(r,t)$
from (\ref{rho}) and (\ref{eq2-2}) and then determine the
Misner-Sharp mass $M(v)$ and the surface radius $R_s$ from
Eqs.(\ref{vaidya-m}) and (\ref{vaidya-r}).

\section{Numerical solutions}

 Without loss of
generality, we choose $a(0)=1$.  In order to make numerical calculation
conveniently, we introduce the dimensionless physical quantities as
follows.
\begin{eqnarray*}
\hspace{-0.5cm}&&k\to kR_0^2,\quad\dot a\to \dot aR_0,\qquad \ddot
a\to \ddot
aR_0^2,\quad r\to r/R_0,\\
\hspace{-0.5cm}&&\dot A\to \dot AR_0,\quad\ddot A \to \ddot AR_0^2,\quad \rho \to \rho R_0^2/{8\pi G},\\
\hspace{-0.5cm}&&p\to pR_0^2/{8\pi G},\qquad q^{\mu}\to q^{\mu}
R_0^2/{8\pi G},
\end{eqnarray*}
where $R_0$ is the initial radius of star.  Note that we have
already chosen the unit of $c=1$.

In the following, we study several examples numerically.

\begin{figure}[t]
\includegraphics[scale=0.72]{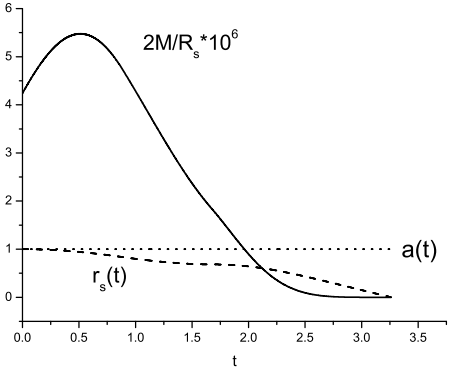}
\caption{Numerical solution of the gravitational collapse of
spherical star with heat flow in the case $p|_{r=0}={\rho}/3|_{r=0}$
and the $u_0^-$ solution in Eq.(\ref{u0}). The horizontal axis is
the time $t$ in the unit of $R_0/c$. In the process of evolution,
$r_s(t)$ (dash curve, in the unit of $R_0$) decreases monotonically
while $a(t)$ (dotted curve) keeps almost a constant.  $2M/R_s$ (real
curve) in the unit of $c^2/G$ increases first and then goes to 0 as
$R_s \to 0$, which implies that the star disappears at the end of
the collapse.}
\end{figure}

\subsection{$p={\rho}/3$ at center}

Suppose that the equation of state at the center of a star take the
radiation form, i.e. $p|_{r=0}={\rho}/3|_{r=0}$, which was also
discussed in Ref. \cite{Harko}.  We are interested in the case that
the surface of a star is almost stationary in the initial state, namely,
\be
&&r_s(t=0)=R_s(t=0)=R_0, \label{ic1} \\
&& \dot R_s(t=0)\approx 0.\label{ir}
\ee
%$.
Therefore, we consider the initial conditions (\ref{ic1}) and
\be
\dot a(t=0)=0, \label{ic2}
\ee

In such a case, only the $u_0^-$ solution in Eq.(\ref{u0}) can reach
(\ref{ir}) from the initial conditions (\ref{ic1}) and (\ref{ic2}).
Fig. 1 presents the numerical solution in the case.
Since $\rho$ and $p$ are determined by Eqs.(\ref{rho}) and
(\ref{eq2-2}), the initial state of the star is not homogeneous and
the equation of state in the star is generally deviated from the
radiation.  In the figure, the time and radius $r_s$ are in the
units of $R_0/c$ and $R_0$, respectively. In the process of
evolution, $r_s(t)$ (dash curve) decreases monotonically while
$a(t)$ (dotted curve) almost keeps a constant.  $2M/R_s$ (real
curve) in the unit of $c^2/G$ increases first and then goes to 0
more quickly than $R_s$ itself as $R_s \to 0$.  It implies that the
star will disappear without the appearance of a horizon in the
process.

\begin{figure}[t]
\includegraphics[scale=0.72]{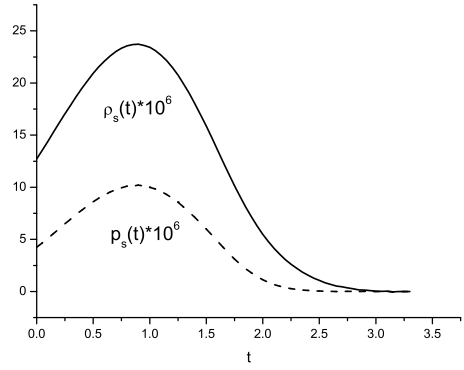}
\caption{Evolution of $\rho$ and $p$ at the boundary.  The
horizontal axis is the time in the unit of $R_0/c$.  $\rho$ and $p$
are in the unit of $c^4/(8\pi G R_0^2)$. }
\end{figure}

Fig. 2 shows the evolution of $\rho$ and $p$ at the boundary, in
unit of $c^4/(8\pi GR_0^2)$.  The emission of the star arrives its
maximum value at about $0.9 R_0/c$ after the beginning of the
collapse. In the late stage of the process, the equation of state at
the boundary as well as at each point in the star tends to $p=\frac
1 3 \rho$.  At the end of the process both the energy density and
the pressure become 0, which confirms that the whole star is
radiated out into space in the process.

Furthermore, our numerical analysis shows that the star with
$p|_{r=0}=\frac 1 3 \rho|_{r=0}$ will radiate its whole mass in the
process of the collapse, without the appearance of a horizon, for
different initial values.

\begin{figure}[b]
\includegraphics[scale=0.72]{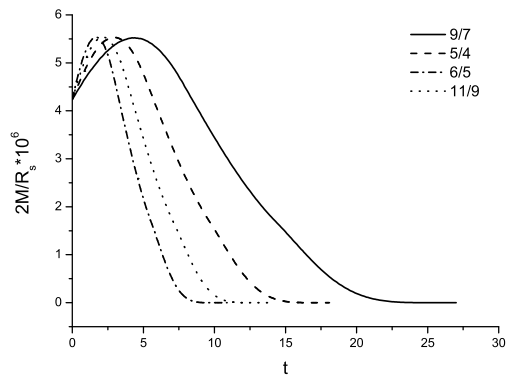} \caption{The ratios of mass 
to radius of collapsing stars for numerical
solutions with $p|_{r=0}={\rho^{\gamma}}|_{r=0}$, and $\gamma=9/7,
5/4, 6/5, 11/9$. $t$ is in the unit of $R_0/c$.}
\end{figure}
\subsection{$p=C\rho^\ga$ at center}

\begin{figure}[t]
\centerline{\includegraphics[scale=0.72]{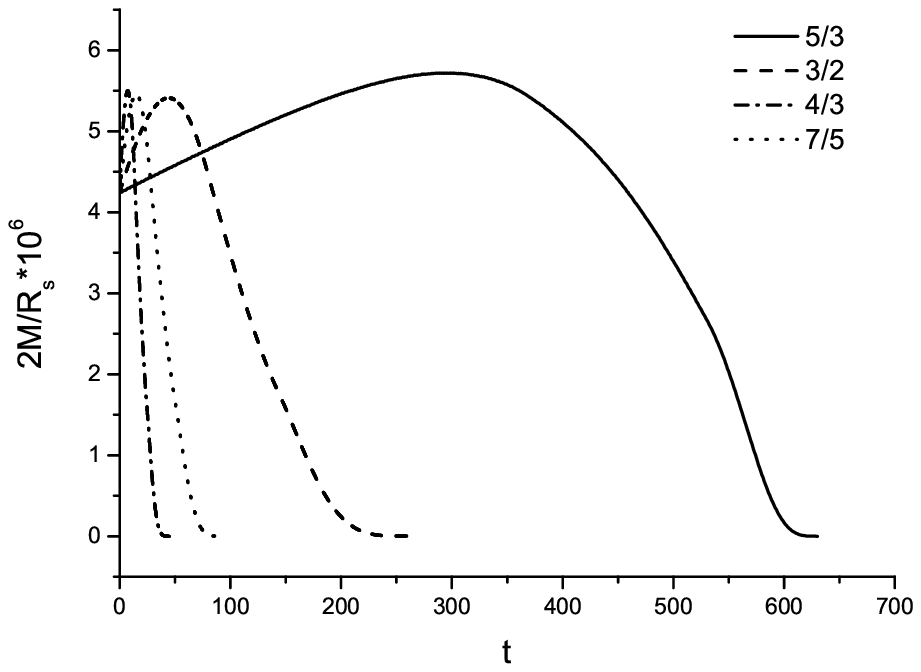}}
\caption{The ratios of mass 
to radius of collapsing stars for numerical
solutions with $p|_{r=0}={\rho^\ga}|_{r=0}$, and
$\gamma=5/3, 3/2, 4/3, 7/5$. $t$ is in the unit of $R_0/c$.}
\end{figure}

Now, let us suppose that the equation of state at the center have
the forms of polytropes.  Again, we only consider the $u_0^-$
solution in Eq.(\ref{u0}) with the initial conditions (\ref{ic1})
and (\ref{ic2}).  Figs. 3 and 4 present the the numerical solutions of
Misner-Sharp mass in the cases $\ga=6/5,~ 11/9,~ 5/4,~ 9/7,~ 4/3,~
7/5,~ 3/2,~5/3$ and $C=1$, respectively.  The time 
is again in the unit of $R_0/c$. The behaviors of $r_s$ for all cases are 
similar to the one in Fig.1.  These
figures show that the evolutions of the stars with different
adiabatic indexes have the similar behavior to that in Fig. 1 but with 
different collapsing time scale.
Namely, the whole star will be emitted out without a 
remnant eventually in all these cases.

\begin{figure}[b]
\centerline{\includegraphics[scale=0.72]{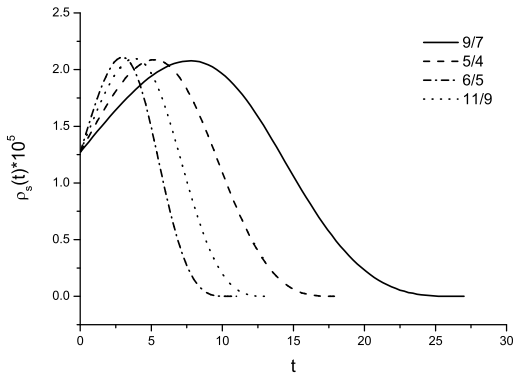}} \caption{Evolution
of $\rho$ at the boundary for the cases of $\gamma=9/7$, $5/4$, $6/5$, and $11/9$. The
horizontal axis is the time in unit of $R_0/c$. $\rho$ is
in the unit of $c^4/(8\pi G R_0^2)$.}
\end{figure}

\begin{figure}[t]
\centerline{\includegraphics[scale=0.72]{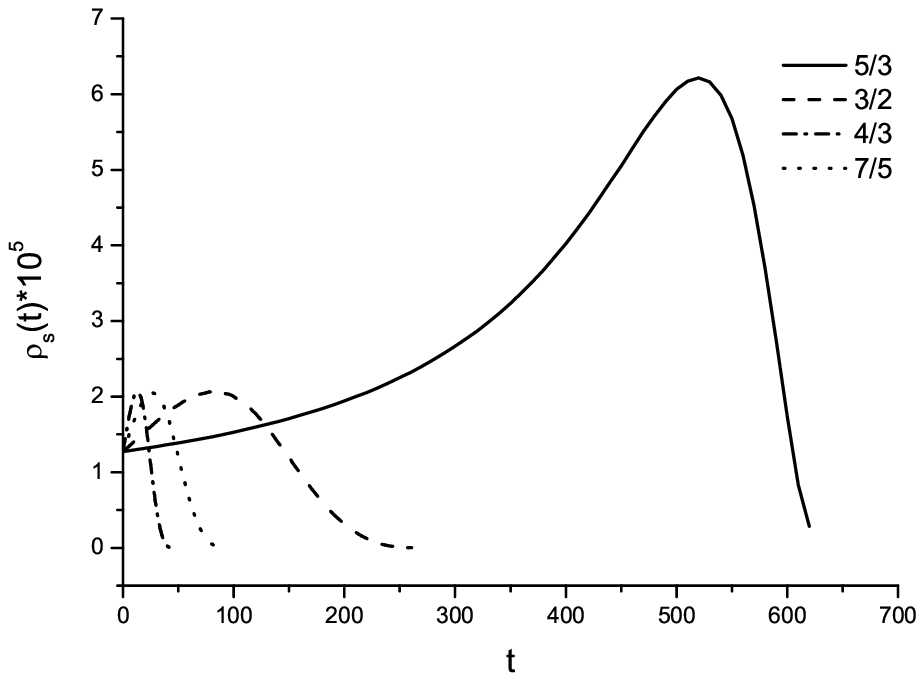}} \caption{Evolution
of $\rho$ at the boundary for the cases of $\gamma=5/3$, $3/2$, $4/3$, and $7/5$.}
\end{figure}
\begin{figure}[b]
\centerline{\includegraphics[scale=0.72]{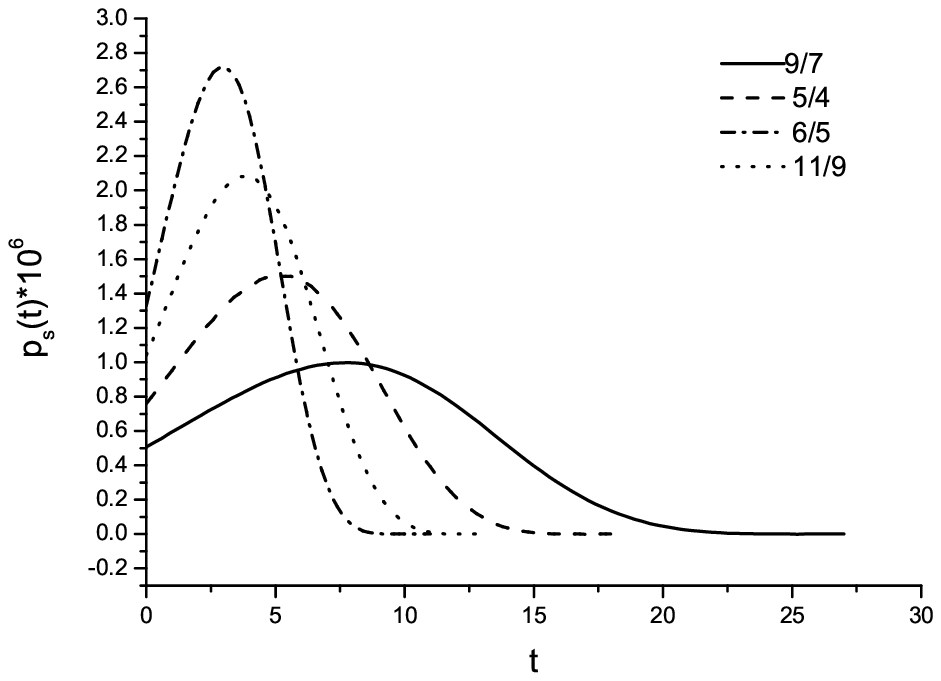}} \caption{Evolution
of $p$ at the boundary for the cases of $\gamma=9/7$, $5/4$, $6/5$, and $11/9$. The
horizontal axis is the time in unit of $R_0/c$.  $p$ is
in the unit of $c^4/(8\pi G R_0^2)$.}
\end{figure}
\begin{figure}[t]
\centerline{\includegraphics[scale=0.72]{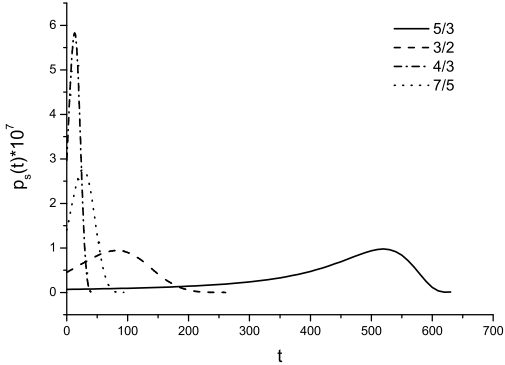}} \caption{Evolution
of $p$ at the boundary for the cases of $\gamma=5/3$, $3/2$, $4/3$, and $7/5$.}
\end{figure}

Figs. 5 and 6 show the evolutions of $\rho$ at the boundary
for $p=\rho^\ga$ with $\gamma=9/7$, $5/4$, $6/5$, $11/9$, $5/3$, 
$3/2$, $4/3$, and $7/5$, respectively, in unit of $c^4/(8\pi GR_0^2)$. 
Figs. 7  and 8 show the evolutions of $p$ at the boundary
for the same parameters.  The times to emit all energy, $T$, are 
shown in the following Table.  The last line in the Table gives the numerical
values for the star with a solar radius initially. \  At the end of the process both\smallskip\\
Table.  \parbox[t]{6.3cm}{Time to emit all energy for the star with  $p=\rho^\ga$ at its center.}\\ 
\begin{tabular}[b]{cccccccccc}
\hline
$\ga$ & 6/5 & 11/9&5/4 & 9/7& 4/3& 7/5&3/2 & 5/3 \\
\hline
$cT/R_0$ &11&14& 18 & 27& 45& 94& 262 & 630\\
$T (s)$   &25.3 &32.7&42.0&62.7&104.5&218.7&608.2&1462.5\\
\hline
\end{tabular}%\smallskip\\ 
  
\noindent the
energy density and the pressure become 0, which confirms that the
whole star is radiated out into space in the process.

\begin{figure}[t]
\centerline{\includegraphics[scale=0.72]{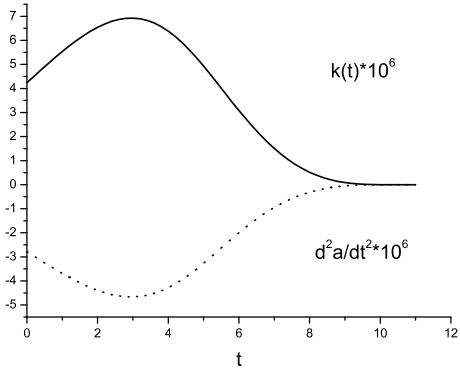}}
\caption{Numerical solution for $k(t)$ and $\ddot{a}(t)$ in the case of
$p|_{r=0}={\rho}^{6/5}|_{r=0}$ and the $u_0^-$ solution of
Eq.(\ref{u0}).}
\end{figure}

Finally, we give the evolution of $k(t)$ and $\ddot a$ in Fig. 9.
Obviously, $k(t)$ and $\ddot a$ have the same order of magnitudes,
which is consistent with the fact that the equation of state at the
center of the star serves as a free input parameter.

\section{Concluding remarks}
We have shown a new approach to study the gravitational collapse of a
spherical, inhomogeneous, fluid star with heat flow.  By use of the Ansatz
of the generalized Friedmann-Robertson-Walker metric, we obtain the new
solutions of Einstein's field equations.  These solutions share the
same properties: the initial mass and radius are finite and whole stars will
be emitted in the process of the gravitational collapse in a finite
interval, without the formation of a horizon, and a Minkowski
spacetime is left at the end of the process.  This is quite
different from the standard evolution of the gravitational collapse
of a star.  In the standard process, a white dwarf, or a neutron
star, or a black hole, or even a singularity will be formed at the final
state and only a part of whole mass will be emitted.  Obviously, the numerical
solutions of the Einstein's field equations presented here are the
description of another type of collapsing evolution beyond of the
standard process.

As mentioned in Introduction, the numerical solutions with the
property that all energy of the star in our model will be emitted in
the process of collapse and an empty flat spacetime will be left
behind have been obtained before \cite{Hirschmann, SG, FST}.
Nevertheless, our solutions are more realistic than the previous
ones. This is partly because in our solutions there is no
self-similarity which will break down the asymptotic flatness of
spacetime and partly because there is no initial singularity and the
initial mass and radius are finite.  Another reason is that the
interiors of stars in nature are inhomogeneous and having the
equation of state of polytrope.  Our solutions presented here
%in the table 
share these characters.  In particular, when we apply our 
solution to a star with about a
solar mass $M_\odot$, a solar radius $R_\odot$ and $p=\rho/3$ at the
center, a huge mount of energy (about $1.8\times 10^{54}$ erg) will
be emitted into space within 7.57 seconds, which are typical values
for a gamma-ray burst.\cite{GRB}  If stars have the same size and
$p=\rho^\ga$ at their centers, the times to emit all energy, $T$, will
range from 25s to 1462.5s.  They are all reasonable values for gamma-ray 
bursts. And according to our numerical solutions we believe that if $\ga$ changes continuously,
then the times to emit all energy will change continuously. Thus, the gravitational collapse of such
a star might provide a new energy mechanism for gamma-ray bursts. Of
course, detailed investigations of applying the solution to describe
gamma-ray bursts are needed.

\omits{ are 25.3
seconds for $\ga =6/5$, 32.7 seconds for $\ga=11/9$, 42.0 seconds
for $\ga=5/4$, 62.7 seconds for $\ga=9/7$, 104.5 seconds for $\ga =
4/3$, 218.7 seconds for $\ga=7/5$, 608.2 seconds for $\ga =3/2$,
1462.5 seconds for $\ga=5/3$ respectively.}

\medskip

We would like to thank Professors H.-Y. Guo and Z.-J. Shang for
helpful discussion. This work is partly supported by the Natural
Science Foundation of China under Grant Nos. 90403023, 10575106 and 10775140
and Knowledge Innovation Funds of CAS (KJCX3-SYW-S03).  One of the authors (C.-B. G.) got the
support in the initial stage of the present work from the
Interdisciplinary Center for Theoretical Study, University of
Science and Technology of China.

\appendix
\section{Boundary conditions on the star surface}

The motion of the star surface can be described by a time-like
three-space $\Sigma$ which divides the spacetime into the interior
$\mathcal{V}^{-}$ and the exterior $\mathcal{V}^{+}$. \omits{As a
boundary, }The induced metric on and the extrinsic curvature of
$\Sigma$ are
\be &&\hspace{-0.7cm}\left. ds^2_{\pm}\right |_{\Sigma} =
\left(g^{\pm}_{\mu\nu}dx^{\mu}_{\pm}dx^{\nu}_{\pm}\right)_{\Sigma},\label{MTT}\\
&&\hspace{-0.7cm}K^{\pm}_{ij}=-n^{\pm}_{\alpha}\frac{\partial^2x^{\alpha}_{\pm}}{\partial\xi^i\partial\xi^j}
               -n^{\pm}_{\alpha}\Gamma^{\pm\alpha}_{\mu\nu}\frac{\partial x^{\mu}_{\pm}}{\partial\xi^i}\frac{\partial x^{\nu}_{\pm}}{\partial\xi^j},
\ee respectively, where $x^{\mu}_{\pm},~ g^{\pm}_{\mu\nu}$ and
$\Gamma^{\pm\alpha}_{\mu\nu}$ are the coordinates, metrics and
affine connections of $\mathcal{V}^{\pm}$ respectively; $\xi^i$ is
the intrinsic coordinate of $\Sigma$ and $n^{\pm}_{\alpha}$ are the
unit covariant vectors normal to $\Sigma$.

Following Israel \cite{Israel66}, the junction conditions are the
continuity of the induced metric and extrinsic curvature, \be
&&\left. ds^2_-\right|_{\Sigma}=\left. ds^2_+\right|_{\Sigma},\label{JC1}\\
&&K^{-}_{ij}=K^{+}_{ij}.\label{JC2} \ee

For the spherically symmetric collapse with heat flow and a
radiating surface, a detailed discussion was given by Santos
\cite{Santos} with the interior
solution in isotropic, comoving coordinates \be
&&\hspace{-1cm}ds^2_{-}=A^2(\tb,\rb)d\tb^2-B^2(\tb,\rb)[d\rb^2+\rb^2d\Omega^2],
\ee and the exterior Vaidya solution \be
&&\hspace{-1cm}ds^2_{+}=(1-\frac{2GM(v)}{R})dv^2+2dv d{R}-R^2 d\Om^2
.\label{Vdy} \ee A similar discussion to Santos' can be taken for
the more general interior solution in comoving coordinates, \be
ds^2_{-}=e^{2\phi}d\tb^2-e^{2\psi}d\rb^2-\rc^2d\Omega^2,\label{Cmv}
\ee where $\phi,~\psi$ and $\rc$ are functions of $\tb$ and $\rb$.

In the comoving coordinates, the equation of motion of the star
surface is $\rb=\rb_{\Sigma} (= \mbox{constant})$. The corresponding
hypersurface equation is $f^{-}_{}(\tb,\rb)=\rb-\rb_{\Sigma}=0$. The
space-like vector $\partial f^{-}_{}/\partial x_{-}^{\alpha}$ is
normal to $\Sigma$ and by normalization
$g^{\mu\nu}_-n^-_{\mu}n^-_{\nu}=-1$ one will have \be
n^{-}_{\alpha}=(0,e^{\psi},0,0). \ee By utilizing the metrics
(\ref{Vdy}) and (\ref{Cmv}) in the first junction condition
(\ref{JC1}), one can have \be
&&\hspace{-1.2cm}R|_{\Sigma}=\rc(\tb,\rb_{\Sigma}),\label{MT1}\\
&&\hspace{-1.2cm}\left[(1-2GM/R)dv^2+2dvdR\right]_{\Sigma}=\left.
e^{2\phi}d\tb^2\right|_{\Sigma}.\label{MT2} \ee The equation of
motion of star surface in the exterior space is given by the above
equations. By supposing the hypersurface equation as
$f^{+}(v,R)=R-R(v)=0$, the normal covariant vector $n^{+}_{\alpha}$
can be written as \be n^{+}_{\alpha}=\lambda(\left.
-\frac{dR}{dv}\right|_{\Sigma},1,0,0). \ee The normalization
$g^{\mu\nu}_+n^+_{\mu}n^+_{\nu}=-1$ gives \be
\lambda^2\left[(1-2GM/R)+2(dR/dv)\right]_{\Sigma}=1. \ee Utilizing
the equation (\ref{MT2}), one easily gets
$\lambda=e^{-\phi}\dot{v}|_{\Sigma}$, and then \be
n^{+}_{\alpha}&=&e^{-\phi}\dot{v}(-dR/dv,1,0,0)|_{\Sigma} \nno \\
                &=&e^{-\phi}(-\dot{R},\dot{v},0,0)|_{\Sigma},
\ee where over-dots denotes the derivative with respect to $\tb$.

The intrinsic coordinates of $\Sigma$ can be conveniently chosen as
$\xi^{i}=(\tb,\theta,\varphi)$. Now with the obtained normal vectors
and metrics, the induced extrinsic curvature can be calculated out.
Their nonzero components are \be
\hspace{-0.4cm}&K^{-}_{\tb\tb}&=\left. -e^{2\phi-\psi}\phi^{\prime}\right|_{\Sigma},\\
\hspace{-0.4cm}&K^{-}_{\th\th}&=\left. e^{-\psi}\rc\rc^{\prime}\right|_{\Sigma},\\
\hspace{-0.4cm}&K^{-}_{\varphi\varphi}&=K^{-}_{\th\th}\sin^2\th,\\
\hspace{-0.4cm}&K^{+}_{\tb\tb}&=\left. e^{\phi}
\left(\frac{\ddot{v}}{\dot{v}}-\dot{\phi}-\frac{GM}{R^2}\dot{v}\right)\right|_{\Sigma},\\
\hspace{-0.4cm}&K^{+}_{\th\th}&=\left. e^{-\phi}R
\left[\dot{R}+(1-\frac{2GM}{R})\dot{v}\right]\right|_{\Sigma},\\
\hspace{-0.4cm}&K^{+}_{\varphi\varphi}&=K^{+}_{\th\th}\sin^2\th. \ee
The second junction condition (\ref{JC2}) (having only two
independent equations because of spherical symmetry) can therefore
be rewritten  as \be \hspace{-0.8cm}\left.
-e^{-\psi}\phi^{\prime}\right|_{\Sigma}&=&
  \left. e^{-\phi}
\left(\frac{\ddot{v}}{\dot{v}}-\dot{\phi}-\frac{GM}{R^2}\dot{v}\right)\right|_{\Sigma},\label{CV1}\\
\hspace{-0.8cm}\left. e^{-\psi}\rc\rc^{\prime}\right|_{\Sigma}&=&
  \left. e^{-\phi}R
  \left[\dot{R}+(1-\frac{2GM}{R})\dot{v}\right]\right|_{\Sigma}.\label{CV2}
\ee With the help of the metric continuity equations (\ref{MT1}) and
(\ref{MT2}), the equation (\ref{CV2}) can be reduced to \be
M(v)&=&\left.\frac{\rc}{2G}\left(1+e^{-2\phi}\dot{\rc}^2-e^{-2\psi}\rc^{\prime 2}\right)\right|_{\Sigma}\nno \\
& \equiv &m(\tb,\rb),\label{CVmass} \ee where $m(\tb,\rb)$ is the
well-known Misner-Sharp mass function\cite{Misner-Sharp}.
Substituting $M(v)$ into the equation (\ref{CV2}), one also gets \be
\dot{v}|_{\Sigma}=\left.
\frac{e^{\phi}}{e^{-\phi}\dot{\rc}+e^{-\psi}\rc^{\prime}}\right|_{\Sigma}.
\ee Substituting $M(v)$ and $\dot{v}$ into equation (\ref{CV1}), one
now gets \be\label{CV11} \hspace{-0.5cm}&&\left.
e^{-\phi-\psi}(\dot{\rc}\phi^{\prime}+\rc^{\prime}\dot{\psi}-
\dot{\rc}^{\prime})\right|_{\Sigma} \nno \\
\hspace{-0.5cm}&=&\left[e^{-2\phi}\left(\ddot{\rc}+\frac{\dot{\rc}^2}{2\rc}  -\dot{\rc}\dot{\phi}\right) \right. \nno \\
\hspace{-0.5cm}&& \left.
 -e^{-2\psi}\left(\frac{\rc^{\prime 2}}{2\rc}+\rc^{\prime}\phi^{\prime}\right)+\frac{1}{2\rc}\right]_{\Sigma}.
\ee

Suppose that the collapse material be a shear-free fluid with heat
flow described by the stress-energy tensor (\ref{stress}).  In the
comoving coordinates, the unit four-velocity vector
$u^{\mu}=(e^{-\phi},0,0,0)$ is normal to the space-like heat flow
vector $q^{\mu}=(0,q^1,0,0)$. By solving the Einstein equation, one
gets \be
\hspace{-1.cm}8\pi{}Gq^{1}&=&-e^{-\phi-2\psi}\frac{2}{\rc}(\dot{\rc}\phi^{\prime}+\rc^{\prime}\dot{\psi}-\dot{\rc}^{\prime}),\label{qsolution}\\
\hspace{-1.cm}8\pi{}Gp&=&e^{-2\phi}\frac{2}{\rc}\left(\dot{\rc}\dot{\phi} -\ddot{\rc}-\frac{\dot{\rc}^2}{\rc^2}\right)\nno \\
&& +e^{-2\psi}\left(\frac{\rc^{\prime
2}}{\rc^2}+\frac{2}{\rc}\rc^{\prime}\phi^{\prime}\right)-\frac{1}{\rc^2}.\label{psolution}
\ee Now from the equations (\ref{CV11}), (\ref{qsolution}) and
(\ref{psolution}) it is easy to see that \be\label{CVqp}
p|_{\Sigma}=%\left.e^{\psi}q^1\right|_{\Sigma}=
\left.\sqrt{|q^{\mu}q_{\mu}|}\right|_{\Sigma}. \ee In the end, one
can conclude that for spherically symmetric collapse with heat flow
but shear free, the boundary conditions are equations (\ref{MTT}),
(\ref{CVmass}) and (\ref{CVqp}), which are all invariant for the
coordinate transformations on time-radial plane. \omits{\blue (The
middle expression of Eq.(\ref{CVqp}) is not invariant under the
coordinate transformations on the time-radial plane!)} Therefore
although the coordinate is not comoving in our model, the above
three boundary equations are still used directly.

%References

\end{document}